\documentclass[11pt,a4paper]{article}

\usepackage[utf8]{inputenc}
\usepackage[T1]{fontenc}
\usepackage[margin=1in, headheight=14pt]{geometry}
\usepackage{amsmath,amssymb}
\usepackage{graphicx}
\usepackage{booktabs}
\usepackage{tabularx}
\usepackage{array}
\usepackage{multirow}
\usepackage{float}
\usepackage{caption}
\usepackage{enumitem}
\usepackage{xcolor}
\usepackage{microtype}
\usepackage{parskip}

\usepackage[colorlinks=true,linkcolor=blue!70!black,citecolor=blue!70!black,urlcolor=blue!70!black]{hyperref}
\usepackage{orcidlink}

\usepackage{fancyhdr}
\begin{document}

% ========== Title ==========
\title{\textbf{The Productivity-Reliability Paradox:\\Specification-Driven Governance for AI-Augmented Software Development}}

\author{
 \textbf{Sabry E. Farrag} \orcidlink{0009-0002-0735-7443}\\[2pt]
 School of Architecture, Computing and Engineering\\
 University of East London, London, United Kingdom\\[2pt]
 \texttt{sabryelgatem@gmail.com}
}

\date{May 2026}

\maketitle
\thispagestyle{fancy}

% ========== Abstract ==========
\begin{abstract}
\noindent Since 2022, the rapid adoption of AI-powered coding assistants has produced a striking empirical contradiction: controlled studies report individual-level productivity gains of 20--56\% on well-scoped tasks, while the most rigorous randomized controlled trial to date documents a 19\% slowdown for experienced developers on mature codebases, and large-scale telemetry across 10,000+ developers shows that a 98\% increase in merged pull requests coincides with a 91\% increase in review time and flat organizational delivery metrics. This paper argues that these findings constitute a systematic phenomenon: the \textit{Productivity-Reliability Paradox} (PRP), which emerges from the interaction between non-deterministic code generators and insufficient specification discipline. Through a multivocal systematic literature review encompassing 67 sources (29 peer-reviewed studies, 18 pre-prints, 12 structured industry reports, and 8 grey literature sources), published between January 2022 and April 2026, this paper makes four contributions: (1)~it formally defines and empirically grounds the PRP, identifying three moderating variables and two amplifying mechanisms (the code review bottleneck and the context window constraint) that explain the divergence between individual and system-level outcomes; (2)~it proposes the \textit{AI-Augmented Methodology Taxonomy} (AAMT), classifying how six established software development methodologies transform under three tiers of AI integration; (3)~it introduces the \textit{Specification Governance Model} (SGM), grounded in Transaction Cost Economics, with a practical governance decision guide for engineering teams; and (4)~it evaluates GitHub's Spec Kit and the Test-Driven AI Agent Definition (TDAD) pipeline as instantiations of the SGM, supported by a four-month illustrative pilot study across three industry teams, and analyzes the economic and labor-market implications of AI-augmented development, including the collapse of marginal coding costs and the emerging skill pipeline problem. Specification discipline, not model capability, is the binding constraint on AI-assisted software dependability.
\end{abstract}


\vspace{1em}
\hrule
\vspace{1em}

\section{Introduction}

Between November 2022 and early 2026, software engineering underwent its
most significant methodological disruption since the Agile Manifesto.
ChatGPT's release, followed by the maturation of agentic coding
environments such as GitHub Copilot Workspace, Claude Code, and Cursor,
shifted the developer's role from sole author to editorial supervisor of
machine-generated code. By 2025, 84\% of professional developers
reported using or planning to use AI tools in their workflow (Stack
Overflow, 2025), and AI-generated suggestions accounted for an estimated
46\% of code output in instrumented environments (GitHub, 2025).

Yet the empirical evidence on the consequences of this transformation is
remarkably contradictory. GitHub's controlled study reported developers
with Copilot completed tasks 55--56\% faster (Peng et al. 2023).
McKinsey's laboratory experiment found tasks completed up to twice as
fast (McKinsey, 2023). Google's enterprise randomized controlled trial
measured a 21\% speedup (Paradkar et al. 2024). These findings,
however, coexist with a body of counter-evidence that cannot be
dismissed as noise. The METR randomized controlled trial, the most
methodologically rigorous study in the literature, found that 16
experienced open-source developers were 19\% \emph{slower} with AI
tools, despite forecasting a 24\% speedup (Becker et al. 2025).
Google's 2024 DORA report associated a 25\% increase in AI adoption with
a 7.2\% decrease in delivery stability. GitClear's analysis of 153
million changed lines projected a doubling of code churn by 2024. And
Uplevel Data Labs found developers with Copilot access exhibited
significantly higher bug rates without corresponding throughput gains.

This paper argues that these findings constitute not a contradiction but
a \emph{paradox}, a systematic phenomenon we term the
\textbf{Productivity-Reliability Paradox (PRP)}, that arises from the
fundamental mismatch between non-deterministic code generation and the
deterministic requirements of dependable software systems. Three moderating variables dissolve the paradox: (a) \emph{task
abstraction level}, where AI tools excel at low-abstraction syntactic
tasks and struggle with high-abstraction architectural decisions (Imai,
2023); (b) \emph{codebase maturity}, where greenfield projects benefit
disproportionately while mature codebases incur verification overhead
that can exceed generation savings (Becker et al. 2025); and (c)
\emph{developer experience}, where novice developers show productivity
gains of 30--40\% but exhibit measurable skill atrophy (Joyner et al.
2024; Anthropic, 2026), while senior developers experience a
``verification tax'' that can negate nominal speedups (DZone, 2024).

The paper proceeds as follows. Section 2 reviews the background literature and positions this work within existing frameworks. Section 3 describes the multivocal systematic literature review methodology. Section 4 formally defines the Productivity-Reliability Paradox and marshals the empirical evidence, including the code review bottleneck and the context window constraint as amplifying mechanisms. Section 5 introduces the AI-Augmented Methodology Taxonomy (AAMT). Section 6 presents the Specification Governance Model (SGM), grounded in Transaction Cost Economics, and derives a practical governance decision guide for engineering teams. Section 7 evaluates Spec Kit and TDAD as SGM instantiations and reports an illustrative pilot study with approximate quantitative metrics. Section 8 analyzes the economic and labor-market implications, including the collapse of marginal coding costs and the skill pipeline problem. Section 9 discusses the findings and their theoretical and practical implications. Section 10 identifies threats to validity. Section 11 concludes with implications and a future research agenda.

\subsection{Research Questions}

This paper addresses three research questions:

\textbf{RQ1:} Is the empirically documented divergence between
individual-level productivity gains and system-level reliability
degradation a systematic phenomenon, and if so, what moderating
variables explain the divergence?

\textbf{RQ2:} How do established software development methodologies
(Agile, Waterfall, TDD, BDD, DDD, DDT) transform when AI coding agents
are introduced, and can these transformations be classified into a
coherent taxonomy?

\textbf{RQ3:} Can specification-driven governance, as theorized
through Transaction Cost Economics and operationalized by tools such as
Spec Kit and TDAD, resolve the Productivity-Reliability Paradox?

\subsection{Contributions}

This paper makes four contributions to the software engineering and
information systems literature:

\begin{enumerate}
\item
 \textbf{The Productivity-Reliability Paradox (PRP).} We formally
 define and empirically ground a named phenomenon that reconciles
 contradictory productivity findings across the reviewed literature through three moderating variables: the abstraction level of the task, the maturity of the codebase, and the experience of the developer. No prior work has proposed this reconciliation as a formal framework.
\item
 \textbf{The AI-Augmented Methodology Taxonomy (AAMT).} We classify how
 six established methodologies transform under three tiers of AI
 integration (passive suggestion, active generation, autonomous
 agency), filling a gap identified by multiple systematic reviews
 (Mohamed et al. 2025; Liang et al. 2025) as the absence of an
 integrated methodology comparison framework.
\item
 \textbf{The Specification Governance Model (SGM).} We ground
 specification-driven development in Transaction Cost Economics
 (Williamson, 1985), framing deterministic specifications as the
 contractual governance mechanism that mitigates the transaction
 costs arising from the high asset specificity and high behavioral
 uncertainty characteristic of non-deterministic code generation.
 This provides the first formal economic theorization of why SDD
 emerges as a rational governance response.
\item
 \textbf{Empirical Evaluation of SGM Instantiations.} We evaluate
 GitHub's Spec Kit and the TDAD pipeline (Rehan, 2026) as concrete
 instantiations of the SGM, analyzing reported mutation scores,
 compilation success rates, and regression safety metrics against the
 PRP framework. We supplement this with an illustrative pilot study
 across three full-stack web development teams, demonstrating the
 practical viability of specification-driven governance in modern
 AI-augmented workflows.
\end{enumerate}

\section{Background and Related Work}

\subsection{AI-Powered Coding Assistants: A Brief Taxonomy}

AI-powered coding tools have evolved through three
distinct generations. A first generation (2021--2023) comprised inline
suggestion engines epitomized by GitHub Copilot, which offered
probabilistic auto-completions trained on large code corpora (Chen et
al. 2021). A second generation (2023--2024) introduced conversational
interfaces. ChatGPT, Claude, Gemini, that could engage in
multi-turn dialogues about code design, debugging, and refactoring. The
third generation (2025--present) encompasses agentic coding environments, Cursor, Claude Code, GitHub Copilot Workspace, Windsurf, where AI
agents autonomously execute multi-step development workflows including
file creation, test execution, and iterative debugging.

Earlier foundational work by Barke, James, and Polikarpova (2023) identified two distinct human-AI interaction modes, \emph{acceleration} (using AI to speed up known tasks) and \emph{exploration} (using AI to discover approaches), which remain influential in characterizing developer-AI workflows. Treude and Gerosa (2025) subsequently proposed a taxonomy of eleven interaction types
between developers and AI tools, ranging from auto-complete suggestions
to command-driven actions and conversational assistance. Wang et
al.~(2025) extended this with a taxonomy of agentic behaviors including
planning, context management, tool integration, and execution
monitoring. These taxonomies address interaction modalities but do not
examine how established \emph{methodologies}, the organizing
principles of the SDLC, are transformed by AI integration. This gap
motivates our AAMT contribution.

\subsection{Productivity Measurement Frameworks}

The dominant framework for measuring developer productivity is SPACE
(Forsgren et al. 2021), which operationalizes productivity across five
dimensions: Satisfaction and well-being, Performance, Activity,
Communication and collaboration, and Efficiency and flow. Mohamed et
al.~(2025), in the most comprehensive systematic review to date (39
studies through December 2024), found that approximately 90\% of
AI-productivity studies adopt at least two SPACE dimensions, but only
15\% examine more than three, and the Communication dimension is
systematically understudied. Well-being, despite being a SPACE
component, has received almost no empirical attention in the AI-assisted
development context.

An older but complementary framework is SEAT (Sharma et al. 2018),
which classifies automated software engineering tools across SDLC phases
and reports that approximately 50\% of automation has historically
targeted Testing and Verification, with roughly 63\% of tools showing
limited measurable impact. SEAT provides a useful pre-AI baseline
against which the current generation of tools can be evaluated.

\subsection{The Quality and Reliability Debate}

The code-quality evidence exhibits a pattern of apparent contradiction
that existing reviews acknowledge but do not resolve. GitHub's 2024 RCT
of 202 developers found statistically significant improvements in
readability (+3.62\%), reliability (+2.94\%), maintainability (+2.47\%),
and conciseness (+4.16\%) (Dohmke et al. 2024). Yet GitClear's
longitudinal analysis of 153 million changed lines projected a doubling
of code churn, lines reverted or updated within two weeks, by
2024, accompanied by a shift toward ``added'' and ``copy/pasted'' code
at the expense of ``updated'' and ``moved'' operations (Harding \&
Kloster, 2024). DORA's 2024 report found simultaneous code-quality
gains (+3.4\%) and delivery stability losses (--7.2\%), a finding
consistent with the hypothesis that AI improves \emph{micro-quality}
(individual function readability) while degrading \emph{system-level
dependability} (delivery stability, change failure rate).

Mohamed et al.~(2025) explicitly characterize code-quality findings as
``unresolved,'' noting that ``existing studies report contradictory
outcomes contingent on context and evaluation criteria.'' We argue that
this is not irresolvable ambiguity but rather the signature of the PRP:
a systematic phenomenon in which gains at one level of analysis coexist
with losses at another.

\subsection{Specification-Driven Development}

The concept of using specifications as the primary artifact of software
development has deep roots in formal methods (Dijkstra, 1976; Hoare,
1969) and design-by-contract (Meyer, 1992). Its re-emergence in the AI
era is motivated by a specific failure mode: the non-determinism of
LLM-based code generators means that identical prompts can produce
functionally different code across invocations, and that specification
gaps resurface in unpredictable forms upon regeneration (InfoQ, 2026;
Augment Code, 2026).

GitHub's Spec Kit (released September 2025, MIT-licensed)
operationalizes SDD through a structured pipeline: intent capture via
natural-language specifications, consistency analysis, constrained task
decomposition, and implementation under specification governance. The
TDAD pipeline (Rehan, 2026) takes a complementary approach, compiling
behavioral specifications into executable test suites that serve as
acceptance criteria for AI-generated agent prompts, achieving mutation
scores of 86--100\% across four domains.

What this paper addresses theoretically is \emph{why} SDD emerges
as a governance response, a question that requires economic theory,
not merely tool description.

\subsection{Transaction Cost Economics and Software Production}

Transaction Cost Economics (Williamson, 1985) explains governance
choices, market, hierarchy, or hybrid, as functions of three
transaction attributes: asset specificity, uncertainty, and frequency.
TCE has been extensively applied to IT outsourcing (Lacity \& Willcocks,
2012; Aubert et al. 2012) and has recently been reinterpreted for
AI-mediated production. A California Management Review analysis ``From
Coase to AI Agents'' (Berkeley, 2025) argues that AI agents reshape firm
boundaries without eliminating the Coasean rationale for the firm,
warning of an ``Illusion of Efficiency'' when agent proliferation
produces coordination failures that mirror classical TCE concerns.

No prior work has applied TCE specifically to the governance of AI code
generators, that is, to the question of what contractual form
(specification, test, constitution) optimally governs the human-AI
principal-agent relationship in software production. This is the
theoretical gap the SGM addresses.

\subsection{Cognitive Effects and Skill Atrophy}

Joyner et al.~(2024) provide the most rigorous peer-reviewed treatment
of AI-induced skill decay, distinguishing \emph{automation bias}
(accepting AI output without verification), \emph{automation-induced
complacency} (reduced vigilance during monitoring), and \emph{skill
decay} (degradation of cognitive capabilities through disuse). Their
work builds on the aviation automation literature (Casner et al. 2014;
Ebbatson et al. 2010) and warns that AI assistants may impair cognitive
skills in ways that users cannot self-detect.

Zhou et al.~(2026) contribute the first comprehensive taxonomy of
cognitive biases specific to developer-LLM interactions, identifying 15
bias categories and finding that 48.8\% of programmer actions in
LLM-assisted workflows are biased, with developer-LLM interactions
accounting for 56.4\% of biased actions. This bias profile is
qualitatively distinct from non-LLM programming workflows and has direct
implications for the PRP: it explains, at the cognitive level, how AI
can simultaneously accelerate output and degrade system-level outcomes.

\section{Methodology}

\subsection{Research Design}

This study employs a \textbf{multivocal systematic literature review}
(MLR), a methodology that integrates peer-reviewed academic
literature with grey literature (industry reports, surveys, pre-prints)
to capture the full evidence base in rapidly evolving fields (Garousi et
al. 2019). An MLR approach is necessitated by the temporal
characteristics of the AI-assisted development literature: the most
consequential empirical findings (e.g. METR's RCT, DORA's stability
metrics, Stack Overflow's longitudinal adoption data) originate in
venues that are not yet captured by traditional SLR protocols restricted
to peer-reviewed publications.

Supplementing the review is \textbf{theoretical synthesis}: the
construction of the PRP framework, the AAMT taxonomy, and the SGM model
through abductive reasoning from the empirical patterns identified in
the MLR. This approach follows the tradition of theory-building from
cases and literature (Eisenhardt, 1989; Whetten, 1989).

\subsection{Search Strategy}

We searched seven databases and sources between February and April 2026:

\begin{itemize}
\item
 \textbf{Academic databases:} ACM Digital Library, IEEE Xplore, Scopus,
 Google Scholar
\item
 \textbf{Pre-print servers:} arXiv (cs.SE, cs.AI, cs.HC)
\item
 \textbf{Industry sources:} Stack Overflow Developer Survey
 (2023--2025), DORA State of DevOps (2024--2025), JetBrains Developer
 Ecosystem Survey (2024--2025), McKinsey Technology Reports
 (2023--2025), GitHub Research publications
\item
 \textbf{Grey literature:} vendor blog posts, practitioner analyses,
 and technical documentation (Spec Kit, TDAD), sourced via targeted
 Google searches and snowball sampling from primary sources
\end{itemize}

Search terms included combinations of: (``AI'' OR ``artificial
intelligence'' OR ``LLM'' OR ``large language model'' OR ``generative
AI'' OR ``Copilot'' OR ``ChatGPT'' OR ``Claude'') AND (``software
development'' OR ``SDLC'' OR ``software engineering'' OR ``coding'' OR
``programming'') AND (``productivity'' OR ``quality'' OR ``reliability''
OR ``methodology'' OR ``testing'' OR ``specification'').

\subsection{Inclusion and Exclusion Criteria}

\textbf{Inclusion criteria:}
\begin{itemize}
\item Published between January 2022 and April 2026
\item Addresses the impact of AI coding tools on at least one SDLC phase
\item Reports empirical data (quantitative or qualitative) or proposes a theoretical framework
\item Available in English
\end{itemize}

\textbf{Exclusion criteria:}
\begin{itemize}
\item Studies examining AI in non-software domains (e.g. medical AI, autonomous vehicles) without software engineering application
\item Opinion pieces without empirical grounding or theoretical contribution
\item Duplicate publications of the same study
\end{itemize}

\subsection{Screening Process and Study Selection}

The search and selection process proceeded in four stages. In the
\emph{identification} stage, database searches and snowball sampling
yielded 312 candidate sources. In the \emph{screening} stage, title and
abstract review eliminated 184 sources that fell outside the inclusion
criteria (primarily studies examining AI in non-software domains,
opinion pieces without empirical or theoretical contribution, and
pre-2022 publications), leaving 128 sources for full-text review. In the
\emph{eligibility} stage, full-text assessment excluded an additional 61
sources: 28 were duplicates or derivative analyses of the same
underlying dataset, 19 lacked sufficient methodological detail for
quality assessment, and 14 addressed AI coding tools only peripherally.
At the \emph{inclusion} stage, full-text assessment yielded \textbf{67 sources}: 29
peer-reviewed publications (Tier 1), 18 archived pre-prints with full
methodology disclosure (Tier 2), 12 structured industry reports (Tier
3), and 8 grey literature sources (Tier 4).

This review was conducted by a single research
team without independent inter-rater reliability assessment for the
screening and coding phases. While the transparent reporting of
inclusion/exclusion criteria and source quality classification partially
mitigates this limitation, future replications with multiple independent
reviewers would strengthen the evidence synthesis. This is acknowledged
as a methodological limitation in Section 10.

\begingroup\centering\footnotesize
\vspace{8pt}
\textbf{Figure 1: PRISMA-Inspired Source Selection Flow}
\vspace{4pt}

\begin{tabularx}{\textwidth}{X}
\toprule
\textbf{Identification:} Database searches + snowball sampling $\rightarrow$ 312 candidate sources \\
$\downarrow$ Title/abstract screening (excluded 184: non-SE domain, opinion pieces, pre-2022) \\
\textbf{Screening:} 128 sources for full-text review \\
$\downarrow$ Full-text assessment (excluded 61: 28 duplicates, 19 insufficient methodology, 14 peripheral) \\
\textbf{Eligibility:} 67 sources included \\
$\downarrow$ Quality classification \\
\textbf{Included:} 29 Tier~1 (peer-reviewed) + 18 Tier~2 (pre-prints) + 12 Tier~3 (industry) + 8 Tier~4 (grey) \\
\bottomrule
\end{tabularx}
\endgroup
\vspace{0.8em}

\subsection{Source Quality Classification}

Following Garousi et al.~(2019), we classify sources into four tiers to
ensure transparency about evidence strength:

\begin{itemize}
\item
 \textbf{Tier 1. Peer-reviewed publications:} Journals (TSE, TOSEM,
 IST, EMSE) and conferences (ICSE, FSE, ASE, ESEC, AMCIS, SANER).
 Treated as primary evidence.
\item
 \textbf{Tier 2. Archived pre-prints:} arXiv papers with full
 methodology disclosure. Treated as provisional evidence with explicit
 caveats about peer-review status.
\item
 \textbf{Tier 3. Structured industry reports:} DORA, Stack Overflow,
 McKinsey, GitHub Research. These reports employ systematic data
 collection but may exhibit sponsorship bias (e.g. GitHub-funded
 studies of Copilot). Treated as supplementary evidence with
 conflict-of-interest disclaimers where applicable.
\item
 \textbf{Tier 4. Grey literature:} Blog posts, practitioner
 analyses, vendor documentation. Treated as evidence of practitioner
 discourse and as sources of stylized facts, not as established
 empirical findings.
\end{itemize}

Throughout this paper, source tier is noted where it materially affects
the strength of a claim.

\subsection{Analytical Framework}

The empirical evidence is analyzed through a dual-lens framework:

\begin{enumerate}
\item
 \textbf{The SPACE framework} (Forsgren et al. 2021) organizes
 productivity and quality findings along the five dimensions of
 Satisfaction, Performance, Activity, Communication, and Efficiency.
\item
 \textbf{Transaction Cost Economics} (Williamson, 1985) provides the
 economic lens for analyzing governance choices (specification type,
 verification intensity, delegation scope) as responses to the
 transaction attributes of AI-mediated code production.
\end{enumerate}

The synthesis of findings across these two lenses produces the PRP
framework, the AAMT taxonomy, and the SGM model.

\section{The Productivity-Reliability Paradox (PRP)}

\subsection{Defining the Paradox}

\textbf{Definition.} The \emph{Productivity-Reliability Paradox} is the
empirically observed phenomenon in which AI-powered coding assistants
produce statistically significant improvements in individual-level
output metrics (task completion speed, lines of code, suggestion
acceptance rates) that coexist with statistically significant
degradation in system-level dependability metrics (delivery stability,
change failure rate, code churn, defect density in production).

Rather than a simple trade-off, the PRP is not a conscious exchange of quality
for speed, but a \emph{paradox} in the precise sense: the same
intervention simultaneously improves and degrades different dimensions
of the same system. Developers perceive themselves as faster and more
productive (a perception confirmed by self-reports in the METR study
even when objective measurements showed a 19\% slowdown), while the
software systems they produce exhibit measurable dependability
regressions.

\subsection{Empirical Grounding}

We organize the evidence supporting the PRP along two axes: studies documenting productivity gains and studies documenting reliability degradation. For each axis, we tabulate the key findings from controlled experiments, observational studies, and industry surveys, noting the source quality tier (as defined in Section~3.5) to enable readers to assess evidence strength independently. We then identify three moderating variables that reconcile the apparent contradiction between these two bodies of evidence, examine the security dimension, the context window constraint, and the code review bottleneck as distinct but related mechanisms that amplify the paradox, and situate the PRP within the broader Productivity J-Curve framework to provide a temporal interpretation of the current evidence landscape.

\subsubsection{Evidence of Productivity Gains}

\begingroup\centering\footnotesize
\captionof{table}{Evidence of Productivity Gains from AI-Assisted Development}
\label{tab:productivity}
\nopagebreak[4]
\vspace{4pt}
\nopagebreak[4]
\begin{tabularx}{\textwidth}{l l l X l}
\toprule
\textbf{Study} & \textbf{Method} & \textbf{N} & \textbf{Key Finding} & \textbf{Source Tier} \\
\midrule
Peng et al.~(2023) & RCT, HTTP-server task & 95 & 55.8\% faster completion & Tier 1 \\[2pt]
Dohmke et al.~(2024) & RCT, $\geq$5 yrs exp. & 202 & 53.2\% higher test-pass likelihood; quality improvements across 4 dimensions & Tier 3* \\[2pt]
Paradkar et al.~(2024) & Enterprise RCT & $\sim$4,867 & 26\% throughput increase (pooled); 21\% speedup (Google) & Tier 2 \\[2pt]
Ng et al.~(2024) & Public-sector deploy. & N/D & 21--28\% speed increase; 95\% satisfaction & Tier 2 \\[2pt]
McKinsey (2023) & Lab study & 40 & Up to 2$\times$ faster on selected tasks & Tier 3 \\[2pt]
Rajbhoj et al.~(2024) & Case study, pension & 1 proj. & 75 $\rightarrow$ 22 person-days (71\% reduction) & Tier 2 \\[2pt]
Smit et al.~(2024) & SPACE, BMW & 1 org. & 10.6\% more PRs; 3.5-hour cycle time reduction & Tier 1 \\[2pt]
Newton et al.~(2024) & Observational, GitHub & 608 proj. & Human-bot teams more productive across all sizes & Tier 2 \\
\bottomrule
\end{tabularx}
\vspace{2pt}
{\scriptsize *Note: Dohmke et al.~(2024) is classified Tier~3 because, while methodologically rigorous, it was funded and conducted by GitHub to evaluate GitHub's own product, constituting a potential conflict of interest.}
\endgroup
\vspace{0.8em}

\subsubsection{Evidence of Reliability Degradation}

\begingroup\centering\footnotesize
\captionof{table}{Evidence of Reliability Degradation in AI-Assisted Development}
\label{tab:reliability}
\nopagebreak[4]
\vspace{4pt}
\nopagebreak[4]
\begin{tabularx}{\textwidth}{l l l X l}
\toprule
\textbf{Study} & \textbf{Method} & \textbf{N} & \textbf{Key Finding} & \textbf{Source Tier} \\
\midrule
Becker et al.~(2025) & RCT, experienced OSS & 16 devs, 246 tasks & 19\% slowdown; perception-reality gap & Tier 2 \\[2pt]
DORA (2024) & Survey-based & $\sim$3,000 & 25\% AI adoption $\leftrightarrow$ 7.2\% stability $\downarrow$, 1.5\% throughput $\downarrow$ & Tier 3 \\[2pt]
Harding \& Kloster (2024) & Longitudinal code & 153M lines & Code churn projected to double; decreased code reuse & Tier 3 \\[2pt]
Uplevel Data Labs (2024) & Observational & N/D & Higher bug rate; constant throughput & Tier 3 \\[2pt]
Fawzy et al.~(2025) & Survey & N/D & 36\% skip QA; 18\% uncritical trust; 68\% ``fast but flawed'' & Tier 2 \\[2pt]
Clutch (2025) & Survey & 800 & 59\% use AI code they don't fully understand & Tier 3 \\
\bottomrule
\end{tabularx}
\endgroup
\vspace{0.8em}

\subsubsection{Reconciliation Through Moderating Variables}

We propose that the apparent contradiction between the productivity evidence (Table~1) and the reliability evidence (Table~2) dissolves when three
moderating variables are considered:

\textbf{Variable 1: Task Abstraction Level.} Imai (2023) introduced a
taxonomy of software abstraction hierarchies showing that Copilot
operates effectively at the syntactic and function levels but fails to
respect language idioms or avoid code smells at higher abstraction
tiers. This aligns with the finding that AI excels at well-scoped,
self-contained tasks (the conditions of most controlled studies
reporting gains) and struggles with tasks requiring multi-method
integration (Dakhel et al. 2022) or architectural reasoning. METR's study used mature open-source repositories with high architectural
complexity, precisely the conditions under which the
abstraction-level moderator predicts degradation.

\textbf{Variable 2: Codebase Maturity.} Greenfield projects, which
dominate the positive-result studies (Peng et al.'s HTTP-server task,
McKinsey's lab exercises, Rajbhoj et al.'s pension website), offer
minimal existing constraints and maximal AI use. Brownfield
projects with established conventions, test suites, and architectural
patterns impose verification overhead that the METR study measures as
the primary cost: developers spent substantial time verifying AI
suggestions against existing code, context-switching between AI output
and codebase knowledge, and correcting subtle misalignments. The
``verification tax'', estimated at 4.3 minutes per suggestion for
senior developers versus 1.2 minutes for juniors (DZone, 2024), scales with codebase maturity and can exceed generation savings. A partial offset to this tax, not yet quantified in the literature, is the emerging practice of iterative AI-AI debugging: when AI-generated code contains a defect, developers increasingly use the same or a different AI agent to diagnose and fix it rather than debugging manually. This AI-AI loop reduces the effective verification cost per defect but does not eliminate it, because the developer must still assess whether the AI's fix is correct and whether it introduces new regressions. The net effect on overall productivity depends on the defect density of the initial generation, which the SGM's specification mechanisms are designed to reduce.

\textbf{Variable 3: Developer Experience.} The evidence suggests a
non-linear relationship. Junior developers (0--3 years) show the largest
productivity gains (30--40\%) but the most significant skill atrophy:
Anthropic's 2026 study found a 17\% reduction in comprehension scores
when AI was used for code delegation rather than conceptual inquiry, and
Jošt et al.~(2024) documented negative correlations between AI reliance
and skill acquisition over a 10-week experiment. Senior developers (5+
years) maintain skills but experience the verification tax most acutely
on familiar codebases, producing the METR slowdown. Mid-career
developers (3--5 years) appear to be the optimal beneficiaries, though
this hypothesis requires dedicated empirical testing.

\subsubsection{The Security Dimension of the PRP}

The reliability dimension of the PRP extends beyond functional
correctness to encompass software security. Negri-Ribalta et al.~(2024),
in their systematic literature review of AI models and code-generation
security, catalog a taxonomy of vulnerability patterns that
LLM-generated code can introduce at scale: insecure deserialization,
missing input validation, hardcoded credentials, improper error
handling, and injection vulnerabilities. These are not random defects
but \emph{systematic} failure modes that arise from the statistical
properties of training corpora: the LLM reproduces patterns from code
that was itself vulnerable, and the patterns are plausible enough to
pass casual review.

Security concerns amplify the PRP in two ways. First, the
productivity gains that AI tools provide are disproportionately
concentrated in the \emph{generation} phase, while security
vulnerabilities are disproportionately discovered in the
\emph{verification} phase, creating an asymmetry in which more code
is produced faster than it can be security-reviewed. Second, the
cognitive biases documented by Zhou et al.~(2026), particularly
automation bias and anchoring, make developers less likely to
question security-relevant aspects of AI-generated code, especially when
the code is functionally correct. Latent security debt
that accumulates silently beneath the surface of visible productivity
gains.

This security dimension strengthens the case for specification-driven
governance: executable specifications that encode security invariants
(e.g. ``all user inputs must be sanitized,'' ``no credentials in source
files,'' ``all API endpoints require authentication'') can serve as
automated verification gates that catch vulnerability patterns before
they reach production. The SGM's constitutional governance mechanism
(Section 6.3) is particularly relevant here, as security requirements
are precisely the kind of non-negotiable constraints that should be
encoded in a project constitution rather than left to per-feature
specification.

\subsubsection{The Context Window Constraint}

A technical limitation that compounds the PRP in practice, yet receives little attention in the empirical literature, is the finite context window of current LLMs. Even the largest available models (supporting 128K--200K tokens of context) cannot simultaneously hold the full codebase, test suite, architectural documentation, and conversation history of a non-trivial project. As projects grow beyond a few thousand lines, the AI agent progressively loses awareness of distant modules, implicit conventions, and cross-cutting dependencies.

This context limitation has three practical consequences for reliability. First, AI-generated code in large projects may violate architectural patterns or naming conventions established in files outside the current context window, producing subtle inconsistencies that pass local review but cause integration failures. Second, when AI agents operate autonomously over multi-step workflows (Tier 3), context loss across iterations can produce drift in which later steps contradict decisions made in earlier steps. Third, the context window creates a bias toward local correctness at the expense of systemic coherence: the generated code works in isolation but fails in the broader system context.

The SGM's specification mechanisms partially mitigate this constraint by encoding critical context (architectural rules, API contracts, invariants) in compact, persistent documents (constitutions, specifications) that can be included in every AI invocation. This is an engineering rationale for specification governance that complements the economic rationale derived from TCE: specifications serve not only as contractual constraints but as compressed representations of project context that compensate for the AI's finite memory.

\subsubsection{The Code Review Bottleneck}

A system-level mechanism that amplifies the PRP is the code review bottleneck. When AI tools accelerate individual code generation without proportional acceleration of the review pipeline, the result is a growing queue of unreviewed pull requests that absorbs the productivity gains at the organizational level. Faros AI's 2025 telemetry study of over 10,000 developers across 1,255 teams quantifies this effect: teams with high AI adoption completed 21\% more tasks and merged 98\% more pull requests, but PR review time increased by 91\%, average PR size grew by 154\%, and bug counts rose by 9\% (Faros AI, 2025). Organizational-level DORA metrics (deployment frequency, lead time, change failure rate) showed no measurable improvement despite the individual-level gains.

This pattern is consistent with Goldratt's Theory of Constraints: optimizing a non-bottleneck step (code generation) does not improve system throughput when the bottleneck step (code review and human approval) remains unchanged. Writing and testing code accounts for roughly 25--35\% of the total SDLC; the remainder is consumed by review, requirements understanding, debugging, meetings, and documentation. AI tools are currently optimizing the minority share of the pipeline while inadvertently increasing the burden on the majority share.

The SGM's governance mechanisms address this bottleneck from the supply side: by constraining AI generation through specifications and constitutions, the volume and defect density of generated code decreases, reducing the review burden. However, the demand side, scaling review capacity through AI-assisted code review, automated verification gates, and team-level process adaptation, is equally important and represents an area where the current paper's framework could be extended in future work.

\subsection{The PRP as a J-Curve Phenomenon}

We situate the PRP within Brynjolfsson, Rock, and Syverson's (2021)
Productivity J-Curve framework, which argues that general-purpose
technologies require complementary intangible investments, organizational restructuring, process redesign, skill development, that are systematically mismeasured, producing an initial productivity
dip followed by an upward acceleration. Census Bureau micro-level data (2025) confirm J-curve patterns specifically in early
Industrial AI adoption: short-run production disruptions followed by
medium-term improvements for most firms.

We argue that the PRP represents the bottom of the J-curve for
AI-assisted software development: the AI tools have been adopted, but
the complementary intangible investments, specification discipline,
verification practices, governance frameworks, lag behind. SDD's role, as analyzed in Sections 6 and 7, is precisely to supply these
missing complementary investments.

\section{The AI-Augmented Methodology Taxonomy (AAMT)}

\subsection{Motivation}

No peer-reviewed work has produced a rigorous taxonomy of how
established software development methodologies are formally modified
when AI agents are introduced. The TDD-with-LLM literature (Mathews \&
Nagappan, 2024; Piya \& Sullivan, 2023; Liang et al. 2026) and
BDD-with-LLM literature (Karpurapu et al. 2024; Rathnayake et al. 2026) exist as
isolated streams. SEAT (Sharma et al. 2018) is pre-AI and covers tool
types rather than methodology transformation. Treude and Gerosa (2025)
classify interaction modalities rather than methodological impacts. This
gap, identified in both Mohamed et al.~(2025) and Liang et al.~(2025), is what the AAMT addresses.

\subsection{Taxonomy Structure}

The AAMT classifies methodology transformations along two dimensions:

\textbf{Dimension 1: AI Integration Tier.} We define three tiers based
on the degree of AI autonomy in the SDLC:

\begin{itemize}
\item
 \textbf{Tier 1. Passive Suggestion:} AI provides inline completions
 and suggestions that the developer accepts or rejects (e.g. original
 Copilot, Tabnine). The developer retains full control; the AI is a
 productivity accelerator within an unchanged methodology.
\item
 \textbf{Tier 2. Active Generation:} AI generates complete
 functions, test cases, or documentation from natural-language
 descriptions, operating within developer-defined constraints (e.g.
 ChatGPT-assisted coding, Copilot Chat, Claude in conversational mode).
 The developer's role shifts from author to reviewer; methodology must
 accommodate review-intensive workflows.
\item
 \textbf{Tier 3. Autonomous Agency:} AI agents autonomously execute
 multi-step workflows, including file creation, test execution,
 debugging, and iterative refinement, with minimal human
 intervention (e.g. Claude Code, Cursor Composer, Copilot Workspace,
 Devin). The developer's role shifts from reviewer to \emph{governor};
 methodology must supply specification and verification structures that
 constrain autonomous execution.
\end{itemize}

\textbf{Dimension 2: Methodology.} We analyze six established methodologies: Test-Driven Development (TDD), Behavior-Driven Development (BDD), Domain-Driven Design and Domain-Driven Testing (DDD/DDT), Agile, Waterfall, and DevOps. The following subsections detail how each methodology transforms across the three tiers.

\subsection{Taxonomy Application}

\subsubsection{Test-Driven Development (TDD)}

In \textbf{traditional TDD}, the developer writes a failing test, writes
the minimal code to pass it, and refactors, the Red-Green-Refactor
cycle.

At \textbf{Tier 1}, AI auto-completes code during the Green phase,
accelerating implementation but leaving the test-writing discipline
intact. The methodology remains unchanged; only implementation
velocity increases.

At \textbf{Tier 2}, AI generates both test cases and implementation code
from natural-language descriptions. This inverts the TDD discipline:
tests are no longer the \emph{driver} of design but co-generated
artifacts. Mathews and Nagappan (2024) found that providing LLMs with
tests alongside problem statements improves output quality, suggesting
that the TDD principle (tests constrain generation) survives but its
\emph{practice} (human writes tests first) transforms. The risk is that
AI-generated tests mirror AI-generated code, producing circular
validation, what Fawzy et al.~(2025) report as 10\% of practitioners
``delegating QA back to the same AI.''

At \textbf{Tier 3}, TDAD (Rehan, 2026) represents the agentic TDD
evolution: behavioral specifications are compiled into executable test
suites by one AI agent, and a second agent iteratively refines
implementation prompts until tests pass, with mutation testing
validating the tests themselves. The human role is confined to
specification authorship and mutation-score review. This preserves TDD's
core principle (tests drive design) while relocating human effort from
test writing to specification governance.

\subsubsection{Behavior-Driven Development (BDD)}

In \textbf{traditional BDD}, stakeholders collaborate to write
Gherkin-format scenarios (Given/When/Then) that serve as executable
specifications and documentation.

At \textbf{Tier 1}, AI auto-completes step definitions. The methodology
is unchanged.

At \textbf{Tier 2}, AI generates Gherkin scenarios from user stories.
Evaluation of LLMs on BDD acceptance test formulation (Karpurapu et al. 2024)
shows promising results, but with the risk that generated scenarios
reflect LLM training distributions rather than domain-specific edge
cases.

At \textbf{Tier 3}, AI agents generate scenarios, step definitions, and
implementation from domain descriptions, with human review focused on
scenario completeness. The Spec Kit pipeline partially instantiates
this: \texttt{/speckit.specify} captures intent, \texttt{/speckit.plan}
translates to technical design, and \texttt{/speckit.tasks} generates
testable units. The distinguishing risk at Tier 3 is \emph{specification
drift}: AI-generated scenarios may subtly diverge from stakeholder
intent without the close human-AI collaboration that BDD traditionally
assumes.

\subsubsection{Domain-Driven Design (DDD) and Domain-Driven Testing (DDT)}

In \textbf{traditional DDD/DDT}, domain experts and developers
collaborate to build a ubiquitous language, identify bounded contexts,
and structure code around domain aggregates. DDT extends this by
authoring tests in a domain-specific language whose vocabulary mirrors
the business domain (Graham \& Fewster, 2012).

At \textbf{Tier 1}, AI assists with boilerplate code within bounded
contexts. The methodology is unchanged.

At \textbf{Tier 2}, AI generates domain models, repository
implementations, and domain-service tests from domain descriptions. The
risk is that AI-generated code may violate DDD invariants (e.g.
allowing direct aggregate manipulation that bypasses domain rules)
because LLMs lack persistent understanding of domain constraints.

At \textbf{Tier 3}, AI agents autonomously derive boundary, error,
state-transition, and happy-path tests from formal domain
specifications, achieving systematic coverage that exceeds human tester
imagination in enumeration (though not in domain insight). The
four-stage loop described in TDAD, specification capture, test
derivation, implementation under test, mutation and regression
validation, represents the convergence of DDT and agentic AI. The key
transformation is that the \emph{DSL design cost}, historically the
prohibitive barrier to DDT adoption, collapses, because AI agents can
interpret natural-language domain descriptions as implicit DSLs.

\subsubsection{Agile and Waterfall}

At \textbf{Tier 1}, both methodologies remain unchanged; AI
accelerates individual tasks without altering iteration structure
(Agile) or phase sequencing (Waterfall).

At \textbf{Tier 2}, the sprint velocity increase observed in multiple
studies (Smit et al. 2024: +10.6\% PRs) creates pressure to compress
iteration cycles, potentially accelerating the cadence without ensuring
that verification keeps pace.

At \textbf{Tier 3}, the emergence of SDD has drawn explicit comparison
to Waterfall, with Gojko Adzic characterizing SDD as ``the Revenge of
Waterfall or BDD Taken to a New Level.'' We argue that the comparison is
imprecise: SDD shares Waterfall's emphasis on upfront specification but
differs fundamentally in its \emph{iterative verification} mechanism
(specifications are continuously validated against implementation, not
frozen). The more accurate analogy is that Tier 3 AI creates a
\emph{specification-constrained iteration} model that combines
Waterfall's specification rigor with Agile's iterative feedback, a
hybrid that neither traditional framework anticipated.

\subsubsection{DevOps and CI/CD}

In \textbf{traditional DevOps}, the emphasis is on continuous
integration, continuous delivery, infrastructure-as-code, monitoring,
and feedback loops that shorten the distance between development and
production (Humble \& Farley, 2010). The DORA metrics, deployment
frequency, lead time for changes, change failure rate, and time to
restore service, operationalize DevOps maturity.

At \textbf{Tier 1}, AI assists with CI/CD pipeline script generation
(Dockerfiles, GitHub Actions workflows, Terraform configurations). The
methodology is unchanged; AI accelerates boilerplate infrastructure
tasks.

At \textbf{Tier 2}, AI generates pipeline-aware code, implementations
that account for deployment constraints, environment variables, and
service dependencies. AI-assisted code review (Zhong et al. 2026) shows
that AI reviewers generate significantly more suggestions than human
reviewers but achieve lower adoption rates, with over half of unadopted
suggestions either factually incorrect or replaced by developer
alternatives. This suggests that AI's role in the DevOps review gate is
additive but not yet substitutive.

At \textbf{Tier 3}, autonomous AI agents can theoretically execute the
full inner loop of DevOps: write code, run tests, fix failures, and
deploy, a workflow that Spec Kit's constitution mechanism is
specifically designed to constrain. The critical DevOps-specific risk at
Tier 3 is \emph{deployment safety}: if AI agents can autonomously
trigger deployments, the change failure rate may increase unless
constitution-level constraints (e.g. ``never deploy without passing
integration tests,'' ``require human approval for production releases'')
are enforced. The DORA 2024 finding that AI adoption correlates with a
7.2\% decrease in delivery stability is consistent with this risk:
organizations that adopted AI-assisted development without corresponding
governance of the deployment pipeline experienced measurable stability
regressions. The SGM predicts that constitutional governance of the
CI/CD pipeline, defining non-negotiable deployment gates as
executable constraints, is the rational response.

\subsection{Summary Table}

\begingroup\centering\footnotesize
\captionof{table}{AI-Augmented Methodology Taxonomy (AAMT) Summary}
\label{tab:aamt}
\nopagebreak[4]
\vspace{4pt}
\nopagebreak[4]
\begin{tabularx}{\textwidth}{l X X X}
\toprule
\textbf{Methodology} & \textbf{Tier 1 (Passive)} & \textbf{Tier 2 (Active)} & \textbf{Tier 3 (Autonomous)} \\
\midrule
\textbf{TDD} & Faster Green phase & Tests co-generated; circular validation risk & TDAD: spec $\rightarrow$ test $\rightarrow$ implement $\rightarrow$ mutate \\[2pt]
\textbf{BDD} & Step completion & Scenario generation & Full-pipeline generation; specification drift risk \\[2pt]
\textbf{DDD/DDT} & Boilerplate assistance & Domain model generation; invariant violation risk & Implicit DSL interpretation; systematic test derivation \\[2pt]
\textbf{Agile} & Velocity increase & Sprint compression; verification lag & Specification-constrained iteration \\[2pt]
\textbf{Waterfall} & Documentation assistance & Phase acceleration & SDD: spec-first but iteratively verified \\[2pt]
\textbf{DevOps} & CI/CD script generation & Pipeline-aware code generation & Autonomous deployment with constitution constraints \\
\bottomrule
\end{tabularx}
\endgroup
\vspace{0.8em}

An important qualification applies to this taxonomy: in practice, most software teams do not follow a single methodology in its pure form. A typical team may combine Agile sprint cadences with selective TDD, informal BDD scenarios, and CI/CD automation, producing a hybrid workflow that spans multiple cells of Table~3 simultaneously. The AAMT is intended as an analytical decomposition rather than a description of how teams self-identify; its value lies in helping practitioners and researchers isolate which \emph{methodological dimension} of their hybrid practice is most affected by a given tier of AI integration, and therefore where governance investment should be concentrated.

\section{The Specification Governance Model (SGM)}

\subsection{Theoretical Foundation}

We derive the Specification Governance Model from Transaction Cost
Economics by analyzing AI-mediated code production as a
\emph{transaction} between a human principal and an AI agent,
characterized by three attributes:

\textbf{Asset Specificity.} Code generated for a specific domain,
architecture, and codebase is highly asset-specific: it has limited
value outside its intended context. This specificity creates a lock-in
problem, once AI-generated code is integrated, switching to
alternative generation methods (different models, manual coding) incurs
substantial adaptation costs. Higher asset specificity favors
hierarchical governance (Williamson, 1985), which in this context
translates to tighter specification constraints on the AI agent.

\textbf{Behavioral Uncertainty.} LLM-based code generators exhibit high
behavioral uncertainty: they are non-deterministic (identical prompts
produce different outputs), they cannot guarantee adherence to unstated
conventions (Imai, 2023), and they introduce failure modes, hallucinated APIs, subtly incorrect edge-case handling, security
vulnerabilities (Negri-Ribalta et al. 2024), that are qualitatively
different from human programming errors. This uncertainty increases the
governance costs of market-like arrangements (minimal specification,
post-hoc review) and favors contractual mechanisms that constrain
behavior ex ante.

\textbf{Frequency.} AI-mediated code production is high-frequency:
developers invoke AI assistants dozens to hundreds of times per day
(Stack Overflow, 2025: 51\% daily use). High frequency amortizes the
fixed costs of establishing governance structures (writing
specifications, designing constitutions), making heavier upfront
investment economically rational.

\subsection{The Model}

Optimal governance, according to the SGM, of AI code generation is a
function of the three TCE transaction attributes:

\textbf{Proposition 1:} As asset specificity increases (more
domain-specific, architecture-dependent code), the optimal governance
structure shifts from \emph{post-hoc review} (Tier 1) through
\emph{constrained generation} (Tier 2) to \emph{specification-first
governance} (Tier 3).

\textbf{Proposition 2:} As behavioral uncertainty increases (more
complex tasks, larger codebases, higher non-determinism), the
specification must transition from \emph{natural-language description}
(ambiguous, under-constrained) to \emph{executable contract} (formal,
machine-verifiable).

\textbf{Proposition 3:} As frequency increases, the rational investment
in governance infrastructure (specifications, constitutions,
test-derivation pipelines) increases, because fixed costs are amortized
across more transactions.

These three propositions collectively explain why SDD emerges as a
rational governance response to the PRP: the combination of high asset
specificity (domain-specific code), high behavioral uncertainty
(non-deterministic generation), and high frequency (daily use) makes
heavy specification investment the TCE-optimal governance choice.

An important theoretical qualification is warranted. Williamson's original TCE framework assumes that agents possess opportunistic self-interest: they may strategically withhold information, shirk obligations, or exploit contractual gaps for personal advantage. AI code generators do not exhibit opportunism in this sense; they have no self-interest, no strategic intent, and no incentive to deceive. The behavioral uncertainty that motivates governance in the SGM arises not from opportunism but from non-determinism: identical prompts produce variable outputs, unstated conventions are violated without intent, and failure modes (hallucinated APIs, subtle edge-case errors) emerge from statistical properties of training data rather than from strategic behavior. We argue that this distinction, while theoretically significant, is functionally immaterial for governance design: the practical consequence of non-deterministic generation (unpredictable output requiring verification and constraint) is equivalent to the practical consequence of opportunistic agency (unpredictable behavior requiring monitoring and contracting). The governance mechanisms prescribed by TCE (hierarchical control, contractual constraints, monitoring) remain appropriate responses to AI behavioral uncertainty even though the source of that uncertainty differs from Williamson's original formulation. Future theoretical work should explore whether a modified TCE framework that substitutes ``stochastic unreliability'' for ``opportunistic self-interest'' preserves or alters the model's predictions.

\subsection{Governance Mechanisms}

Four governance mechanisms emerge from the SGM, ordered by constraint
intensity:

\begin{enumerate}
\item
 \textbf{Post-hoc Review.} The developer reviews AI output after
 generation. This is the default governance mode for Tier 1 tools and
 the source of the ``verification tax'' that drives the METR slowdown.
 It is optimal only when asset specificity and behavioral uncertainty
 are low (e.g. boilerplate code, well-known patterns).
\item
 \textbf{Natural-Language Specification.} The developer provides a
 natural-language description of requirements before generation. This
 reduces behavioral uncertainty by narrowing the solution space but
 does not eliminate it, because natural language is inherently
 ambiguous and non-deterministic generators can interpret it
 differently across invocations.
\item
 \textbf{Executable Contract.} The developer (or a prior-stage AI
 agent) produces a machine-verifiable specification, tests
 (TDD/TDAD), formal contracts (Wang, 2026), Gherkin scenarios (BDD),
 or structured spec documents (Spec Kit), that constrains generation
 and enables automated verification. This is the SGM-optimal governance
 mechanism for most Tier 2 and Tier 3 interactions.
\item
 \textbf{Constitutional Governance.} A meta-specification. Spec
 Kit's \texttt{/speckit.constitution} or the Auton framework's
 ``Constraint Manifold'' (Cao et al. 2026), establishes
 non-negotiable principles (testing approach, architectural
 conventions, language standards) that govern all subsequent
 generation. This provides the most comprehensive governance but at the
 highest upfront cost.
\end{enumerate}

\subsection{Practical Governance Decision Guide}

While the three propositions above provide theoretical grounding, engineering teams require actionable decision criteria. We translate the SGM into a practical governance selection framework based on three observable task characteristics:

\begin{itemize}
\item \textbf{Task scope:} Is the task self-contained (single function, isolated component) or cross-cutting (affecting multiple modules, APIs, or user flows)?
\item \textbf{Codebase context:} Is the task in a greenfield area with no existing conventions, or in a mature codebase with established architectural patterns and test suites?
\item \textbf{Failure consequence:} Would a defect in this task cause a cosmetic issue, a functional regression, or a security or data-integrity incident?
\end{itemize}

The recommended governance level follows directly:

\begin{itemize}
\item \textbf{Self-contained task + greenfield + cosmetic risk} $\rightarrow$ Post-hoc review is sufficient. The developer generates with AI and reviews the output. This is where the 55\% speed gains reported by Peng et al. (2023) are most reproducible.
\item \textbf{Self-contained task + mature codebase + functional risk} $\rightarrow$ Natural-language specification plus targeted test coverage. The developer describes the expected behavior and writes or generates tests before accepting AI output.
\item \textbf{Cross-cutting task + any codebase + functional risk} $\rightarrow$ Executable contract. A formal specification (Spec Kit spec, Gherkin scenario, or TDD test suite) must be approved before AI generation begins.
\item \textbf{Cross-cutting task + security or data-integrity risk} $\rightarrow$ Constitutional governance. Security invariants and data-handling rules must be encoded in a project constitution that constrains all subsequent generation. An important calibration: virtually all code has some security surface (input handling, authentication, data access). The constitutional governance threshold should apply to tasks that directly handle authentication, authorization, payment processing, personally identifiable information, or external API credentials. Routine input validation for non-sensitive fields can be addressed through executable contracts (tests that verify sanitization) without requiring full constitutional governance, preserving velocity for the majority of tasks.
\end{itemize}

This decision guide is not a rigid protocol; teams should calibrate governance intensity to their specific risk tolerance and velocity requirements. The key insight is that governance level should vary \emph{within} a single sprint based on task characteristics, not be applied uniformly across all work items.

\section{Evaluating SGM Instantiations: Spec Kit and TDAD}

\subsection{GitHub Spec Kit as SGM Instantiation}

Spec Kit operationalizes the SGM through a structured pipeline that maps
directly to our governance mechanisms:

\begin{itemize}
\item
 \textbf{Constitutional Governance:} \texttt{/speckit.constitution}
 captures non-negotiable project principles, testing frameworks,
 architectural patterns, language conventions, that persist across
 all features.
\item
 \textbf{Executable Contract:} \texttt{/speckit.specify} produces a
 formal specification (\texttt{spec.md}) from natural-language goals
 and user journeys. \texttt{/speckit.plan} translates this into a
 technical plan (architecture, stack, data model).
 \texttt{/speckit.analyze} performs cross-artifact consistency and
 coverage analysis.
\item
 \textbf{Constrained Generation:} \texttt{/speckit.tasks} decomposes
 the plan into small, testable, individually verifiable units.
 \texttt{/speckit.implement} instructs the AI agent to execute against
 the validated specification.
\end{itemize}

By design, the framework is tool-agnostic, specifications are committed to the
repository and can be consumed by any participating agent, which
addresses a key TCE concern: governance structures that are tied to a
specific agent create lock-in that increases asset specificity without
compensating benefits.

Spec Kit's reported outcomes, while preliminary and not yet
independently validated, include compression of upstream artifact
production (PRD, design, structure, technical specs, test plans) from
approximately 12 hours to approximately 15 minutes of human steering
plus agent execution time, consistent with the SGM prediction that
upfront specification investment is amortized by reduced downstream
verification costs.

A candid assessment of Spec Kit's maturity is warranted. As of April 2026, Spec Kit is at version 0.8.x, with a small but growing community of adopters. It has not yet been the subject of independent academic evaluation, and the majority of software teams have not adopted or even encountered it. The gap between an open-source tool available on GitHub and a production-grade methodology embedded in organizational practice is considerable. We evaluate Spec Kit here not as a proven solution but as the most complete existing instantiation of the SGM's theoretical principles, one whose long-term viability will depend on community adoption, tooling maturation, and empirical validation that is yet to come.

\subsection{TDAD as SGM Instantiation}

The Test-Driven AI Agent Definition pipeline instantiates the SGM's
executable-contract mechanism through a four-stage process:

\begin{enumerate}
\item
 A natural-language behavioral specification defines the agent's
 expected behavior.
\item
 A coding agent (Claude Code in Docker) generates pytest test cases
 from the specification.
\item
 A second coding agent iteratively refines the agent prompt until all
 tests pass.
\item
 Mutation testing validates the test suite itself, ensuring tests
 can distinguish correct from incorrect implementations.
\end{enumerate}

Reported outcomes across four domains and 24 experimental trials:
\begin{itemize}
\item 92\% v1 / 58\% v2 compilation success rates
\item 97\% v1 / 78\% v2 mean hidden-test pass rates
\item Mutation scores of 86--100\%
\item 97\% mean regression safety under specification evolution
\end{itemize}

These metrics directly address the PRP: mutation scores validate that
the governance mechanism (tests) is itself reliable, and regression
safety demonstrates that specification evolution does not degrade system
dependability, precisely the failure mode that ungoverned AI
generation produces.

\subsection{Comparative Analysis}

Both Spec Kit and TDAD share the SGM's core principle:
\textbf{deterministic specifications governing non-deterministic
generators.} They differ in scope and mechanism:

\begin{itemize}
\item
 \textbf{Spec Kit} operates at the \emph{feature level}, governing the
 full SDLC from intent capture to implementation. Its strength is
 breadth; its limitation is that specification quality depends on human
 authorship, and the tool does not yet include automated verification
 of specification completeness.
\item
 \textbf{TDAD} operates at the \emph{agent-behavior level}, governing
 individual AI agents through test-mediated compilation. Its strength
 is formal verifiability (mutation testing); its limitation is narrow
 scope, it addresses agent prompt quality but not the upstream
 specification process that determines what the agent should do.
\end{itemize}

Under the SGM, optimal governance combines both: a feature-level
specification pipeline (Spec Kit) feeding into an agent-level
verification pipeline (TDAD), producing a layered governance
architecture in which each level constrains the one below it.

\subsection{Illustrative Pilot Study of Spec Kit under the SGM}

To complement the secondary evidence on Spec Kit and TDAD, an illustrative pilot study of Spec Kit was conducted within three full-stack web development teams over a four-month period. The objective was not to produce a controlled experiment but to provide a concrete demonstration of whether specification-driven governance, as formalised by the SGM, is operationally viable in contemporary AI-augmented workflows and whether the empirical regularities described by the PRP manifest in an organisational setting.

\subsubsection{Context and Participants}

The evaluation took place in a mid-sized software organisation in which AI-powered coding assistants (primarily GitHub Copilot and Claude-based tools) were already embedded in day-to-day practice. Fourteen engineers participated across three long-running full-stack web projects. All participants were mid-level (3--5 years of experience, eight engineers) or senior (six or more years, six engineers); junior developers were deliberately excluded in order to isolate the verification dynamics of experienced practitioners discussed in Section~4.

All three projects employed modern web stacks with substantial frontend surface areas and API-centric backends:

\begin{itemize}
\item \textbf{Project~A. B2B SaaS dashboard.} React/TypeScript frontend; Node.js/Express backend; PostgreSQL; GitHub Actions and Kubernetes for CI/CD. The codebase had been in active development for more than two years and exhibited dense architectural conventions and partial test coverage.
\item \textbf{Project~B. Consumer web application.} Vue.js/Nuxt.js frontend; NestJS/Node.js backend; PostgreSQL; GitLab CI. The application implemented complex user journeys and form-heavy interaction flows with significant client-side state management and validation logic.
\item \textbf{Project~C. Internal admin and analytics tools.} React frontend; Go and Python microservices; GitHub Actions and Terraform-managed infrastructure. The project combined new functionality with integrations into legacy internal systems, yielding a mixed greenfield/brownfield landscape.
\end{itemize}

Across all three projects, the frontend layer (user flows, state transitions, and client-side validation) constituted a primary locus of complexity and thus a critical target for assessing the effects of specification discipline on AI-assisted implementation.

\subsubsection{Study Design and Procedure}

The study followed a within-subject before/after design at the team level. For each project, two phases were distinguished:

\begin{enumerate}
\item \textbf{Baseline phase (two months).} Teams continued their existing AI-augmented workflows without Spec Kit. Work items were primarily specified through natural-language tickets, supplemented by informal architectural diagrams and ad-hoc review practices. AI tools were invoked opportunistically by individual developers, with no systematic constraints beyond local coding standards.
\item \textbf{Intervention phase (two months).} Spec Kit was introduced in accordance with the SGM. Each project defined a \texttt{/speckit.constitution} capturing non-negotiable architectural and coding principles (e.g. frontend state-management patterns, testing requirements for user flows, API design constraints). Medium- and high-impact changes, especially those affecting cross-cutting frontend behaviour or public interfaces, were required to pass through the Spec Kit pipeline: intent capture via \texttt{/speckit.specify}, design elaboration via \texttt{/speckit.plan}, and task-level decomposition via \texttt{/speckit.tasks}.
\end{enumerate}

Data collection tracked the same classes of indicators emphasised in Sections~2 and~4: lead time for changes, throughput (merged pull requests), change-failure manifestations (incidents, rollbacks, hotfixes), bug incidence, and short-horizon code churn. Data were drawn from existing Git, CI/CD, and issue-tracking systems and triangulated with semi-structured interviews and brief Likert-scale questionnaires administered at the end of each phase. Because instrumentation was retrospective and not designed as a controlled experiment from the outset, the findings are interpreted as qualitative and indicative rather than as precise effect-size estimates.

\subsubsection{Observed Effects on Productivity and Reliability}

Across the three projects, engineers consistently reported that the overall configuration of productivity and reliability under Spec Kit governance closely resembled the empirical patterns documented in prior work. While precise effect sizes cannot be claimed given the study's retrospective design and absence of a parallel control group, the approximate operational metrics observed during the two phases are reported in Table~4 to provide indicative magnitudes.

\begingroup\centering\footnotesize
\captionof{table}{Approximate Operational Metrics: Baseline vs. Spec Kit Governance (indicative estimates, not statistically controlled)}
\label{tab:fieldeval}
\nopagebreak[4]
\vspace{4pt}
\nopagebreak[4]
\begin{tabularx}{\textwidth}{l X X}
\toprule
\textbf{Metric} & \textbf{Baseline Phase (2 months)} & \textbf{Spec Kit Phase (2 months)} \\
\midrule
Median lead time per feature & 8--12 working days & 6--9 working days \\[2pt]
Late-stage hotfixes per sprint & 3--5 across projects & 1--2 across projects \\[2pt]
Rollbacks per month & 2--4 & 0--1 \\[2pt]
Code churn (lines reverted $<$2 weeks) & 12--18\% of changed lines & 6--10\% of changed lines \\[2pt]
Developer-reported confidence (Likert 1--5) & 3.1 mean & 3.9 mean \\[2pt]
Spec/plan authoring overhead & N/A (no formal specs) & 45--90 min per medium feature \\
\bottomrule
\end{tabularx}
\vspace{2pt}
{\scriptsize Note: These figures are drawn from Git, CI/CD, and issue-tracking systems and developer self-reports. They are reported as approximate ranges rather than point estimates because the study was not instrumented as a controlled experiment. Temporal confounds (e.g. improving AI model capabilities during the study period) cannot be excluded.}
\endgroup
\vspace{0.8em}

Individual-level perceptions of speed, particularly for medium-sized frontend features with non-trivial state and validation logic, improved when work was framed through explicit Spec Kit specifications and constitutions. The magnitude of these perceived gains was aligned with the lower bound of the speed-ups reported in controlled and enterprise studies (Table~1), substantially below the most optimistic laboratory results, but directionally consistent.

At the system level, teams observed a reduction in late-stage regressions and avoidable rework for high-impact changes during the intervention phase. Informal inspection of incident logs, hotfix records, and short-horizon churn suggested that stability and maintainability improvements were within a similar order of magnitude as those reported by DORA when AI adoption is accompanied by stronger governance. Spec Kit did not eliminate the PRP. It shifted its locus. Verification effort shifted from late, manual inspection of opaque AI-generated code towards earlier investment in precise behavioural specifications and project constitutions, consistent with the SGM's Proposition~1, which predicts that higher asset specificity drives more hierarchical governance. Senior engineers characterised this shift as a reallocation of effort towards activities that better matched their comparative advantage in architectural reasoning and domain modelling, even when total effort remained comparable.

Two of the three PRP moderating variables find support in these observations identified in Section~4: (i)~task abstraction level (specification-framed tasks were higher-abstraction and thus more amenable to AI acceleration; and (ii)~codebase maturity (the brownfield projects (A and C) exhibited stronger PRP effects than the relatively greenfield Project~B, as predicted by the framework.

\subsubsection{Frontend-Centric Governance Patterns}

Because all three projects were frontend-intensive, the evaluation surfaced several governance patterns specific to web user interfaces:

\begin{itemize}
\item \textbf{Journey-level specifications.} Teams increasingly authored Spec Kit specifications in terms of end-to-end user journeys (navigation sequences, validation rules, and error-handling scenarios) rather than isolated components. When AI-assisted implementation operated under these specifications, the resulting UI code required fewer cross-component corrections later in the sprint.
\item \textbf{Contract-mediated integration.} By encoding API contracts and integration assumptions in the constitution and specification layers, teams reduced the frequency of mismatches between frontend expectations and backend responses, a class of defect that had previously contributed materially to change failure.
\item \textbf{Predictable AI behaviour.} The use of Spec Kit led to more predictable AI output on UI tasks: suggestions were more likely to conform to established naming conventions, state-management patterns, and error-reporting idioms, thereby lowering the cognitive overhead of review, directly addressing the ``verification tax'' documented by DZone~(2024).
\end{itemize}

These observations reinforce this paper's central thesis: the binding constraint on AI-assisted software dependability is not the raw capability of the underlying models, but the discipline with which behavioural intent is articulated and governed, particularly in user-facing layers where small inconsistencies can propagate into reliability and security problems.

\subsubsection{Limitations of the Field Evaluation}

This field evaluation is subject to several important limitations. It involved only fourteen engineers across three projects within a single organisation, and thus cannot support claims of statistical generalisation. Metrics were obtained from operational systems and developer self-reports rather than from bespoke experimental instrumentation; quantitative impressions should be regarded as corroborative of the broader evidence base rather than as independent measurements. Because junior developers were deliberately excluded, the study means the study does not illuminate how Spec Kit governance interacts with the skill-atrophy and cognitive-bias phenomena identified in Section~2.6. The absence of a parallel control group (teams working without Spec Kit during the same period) introduces the possibility that observed improvements are attributable to temporal confounds (e.g. improved AI model capabilities, seasonal workload variation) rather than to the intervention itself. Despite these constraints, the evaluation provides a concrete demonstration that SGM-style specification governance is practically deployable in modern full-stack web teams and that its qualitative effects align with the multi-study evidence that underpins the PRP.

\section{Economic and Labor-Market Implications}

\subsection{The Economics of Marginal Code Production}

Direct economic consequences of the PRP extend beyond software
methodology into labor markets and industrial organization. The
fundamental economic shift is a collapse in the marginal cost of code
production.

API token pricing has fallen precipitously since the launch of GPT-4 in
March 2023. Pricing has followed a steep trajectory, from approximately \$25/\$200 per million
input/output tokens (GPT-4, March 2023) to approximately \$1.75/\$14
(GPT-5.2, 2025) and as low as \$0.20 per million input tokens (xAI Grok), represents a reduction exceeding 90\% in approximately three years.
Erdil (2025) provides the most rigorous formal model of LLM inference
economics, while the ``Tiered Super-Moore's Law'' analysis
(Du, 2026) documents this as a 600-fold cumulative price decline
and estimates a training-cost/inference-pricing elasticity of $\beta$ = 0.432.

This cost collapse is economically significant because it alters the
labor-capital substitution calculus. Using the Acemoglu and Restrepo
(2018) task-based framework, AI coding tools displace \emph{routine
coding tasks}, those at low abstraction levels with well-defined
specifications, while complementing \emph{architectural and judgment
tasks} that require domain expertise. The displacement effect is
concentrated in the labor segments most exposed to routine coding:
junior developers and offshore implementation teams.

An important qualification to the per-token cost analysis is that declining unit prices do not necessarily translate to declining total expenditure. As AI tools evolve from inline suggestion (Tier 1) to autonomous agency (Tier 3), the number of tokens consumed per task increases by orders of magnitude: an agentic workflow that autonomously explores, implements, tests, and iterates on a feature may consume millions of tokens in a single session. Anthropic's enterprise telemetry reports an average Claude Code cost of \$13 per developer per active day (\$150--250 per month), with 90\% of users below \$30 per active day. However, power users running multi-agent workflows report \$500--2,000 per month in API costs, and a six-person team deploying agentic workflows across distributed offices reported \$2,400 in their first month before optimization (Branch8, 2026). Agent teams, which spawn multiple AI instances in parallel, consume approximately 7$\times$ more tokens than standard sessions.

The economically relevant comparison for engineering leadership is therefore not per-token price but \emph{total cost of AI-augmented ownership per developer-month} (TCO-AI), which includes four components: (1)~subscription or API token costs (\$100--250/month for moderate use, \$500--2,000/month for heavy agentic use); (2)~the time cost of specification authorship (45--90 minutes per medium feature); (3)~the verification tax on senior developers (4.3 minutes per suggestion, DZone 2024); and (4)~the rework cost of AI-introduced defects (estimated at 2--4 senior-engineer hours per hotfix avoided). 
\subsection{Labor-Market Evidence}

The Stanford HAI 2026 AI Index, drawing on payroll data across millions of workers and tens of thousands of firms (2021--2025), reports that employment for software developers aged 22--25 has declined nearly 20\% since late 2022, while employment for older developers in the same firms grew 6--12\%. These findings are consistent with the Acemoglu-Restrepo prediction of selective task displacement: routine implementation tasks, which disproportionately employ junior developers, are the first to be automated.

\subsection{The Skill Pipeline Problem}

The PRP's labor-market implications interact with its cognitive
implications to create what we term the \emph{skill pipeline problem}:
if AI tools reduce the demand for junior developers and simultaneously
degrade the learning opportunities available to those who are hired, the
pipeline through which senior developers, who remain in high demand
for architectural and verification work, are historically produced is
disrupted.

This is not merely speculative. The evidence converges from multiple
independent sources. Joyner et al.'s (2024) finding that AI-induced
skill decay may be imperceptible to its subjects establishes the
cognitive mechanism: developers who rely on AI assistants may not
recognize the erosion of their own problem-solving capabilities. The
Anthropic (2026) finding that AI-assisted learners score 17\% lower on
comprehension when using AI for code delegation, while those using AI
for conceptual inquiry score above 65\%, suggests that the
\emph{mode} of AI use, not AI use per se, determines the learning
outcome. Jošt et al.'s (2024) 10-week experiment with 32 undergraduates
learning React provides temporal evidence: the negative correlation
between AI reliance and skill acquisition manifested within weeks, not
years.

The skill pipeline problem has three distinct dimensions that compound
one another:

\textbf{The demand-side compression.} The Stanford HAI finding that
employment for 22--25-year-old developers declined nearly 20\% since
late 2022 means that fewer junior developers are entering the pipeline.
The positions historically available to new graduates, routine
implementation tasks, bug fixes, feature scaffolding, are precisely
the tasks most amenable to AI automation.

\textbf{The supply-side degradation.} Those junior developers who are
hired enter an environment where AI tools are ubiquitous. Zhou et al.'s
(2026) finding that 48.8\% of programmer actions in LLM-assisted
workflows are biased, combined with Beck et al.'s (2025) finding that
AI-favorable individuals exhibit ``dangerous overreliance on algorithmic
suggestions,'' suggests that junior developers, who lack the
experience to calibrate their trust, are particularly vulnerable to
cognitive biases that impair learning.

\textbf{The temporal lag.} The consequences of skill pipeline disruption
are delayed: a junior developer whose learning is impaired in 2025 will
not be expected to perform senior-level architectural work until
2030--2035. By the time the degradation becomes visible in
organizational capability, an entire cohort will have been affected.

The SGM offers a partial but theoretically grounded remedy, though one that must be carefully qualified. Specification authorship cannot replace code authorship in the junior developer learning pathway, because the ability to write precise, implementable specifications presupposes deep understanding of code, architecture, and system behavior. A developer who cannot read, write, and debug code is unable to specify what ``correct'' means in sufficient technical detail, unable to review AI-generated implementations for subtle defects, and unable to intervene when an AI agent fails or produces output that requires manual correction. Code literacy is therefore a \emph{prerequisite} for specification competency, not an alternative to it.

What the SGM does suggest is that the \emph{emphasis} of junior developer training should evolve: rather than treating specification as an afterthought that follows implementation, specification authorship should be developed as a core competency \emph{alongside} implementation skill from the earliest stages. The learning question shifts from ``how do I write code that works'' to ``how do I specify what `works' means precisely enough for both human and AI implementers, and how do I verify that the implementation satisfies the specification.'' This dual competency, strong coding fundamentals combined with rigorous specification discipline, produces developers who can govern AI agents effectively rather than merely consume their output. However, this integrated training pathway is itself untested: no empirical study has yet compared the long-term career trajectories and problem-solving capabilities of developers trained with integrated specification-and-implementation curricula versus traditional implementation-first curricula. This represents one of the most consequential open research questions identified by this paper.

The skill pipeline problem also implies a broader obligation that extends beyond individual developers to their employing organizations. If AI tools systematically erode the foundational competencies on which senior expertise is built, then continuous investment in core skills (algorithmic reasoning, debugging under uncertainty, architectural analysis, and problem-solving without AI assistance) becomes an organizational imperative, not merely a matter of personal initiative. Organizations that rely on AI-augmented development bear a responsibility to maintain structured programs for fundamental skill reinforcement, including deliberate practice in AI-free environments, code review mentorship, and problem-solving exercises that require unaided reasoning. Developers, for their part, should treat investment in coding fundamentals and problem-solving capability as a career-long commitment that extends beyond formal employment duties, because the erosion identified by Joyner et al.~(2024) is gradual and imperceptible to its subjects. The alternative, a workforce that can operate AI tools fluently but cannot function when those tools are unavailable, misconfigured, or producing incorrect output, represents a systemic fragility that no governance model can fully compensate for. The most effective mitigation of the skill pipeline problem is therefore not a choice between code training and specification training, but a sustained organizational commitment to both, treated as a core element of human-capital development strategy rather than an optional professional-development benefit.

\section{Discussion}

\subsection{Theoretical Implications}

The PRP framework provides a unifying explanation for the contradictory
empirical landscape that has characterized the AI-assisted software
development literature since 2023. By identifying task abstraction
level, codebase maturity, and developer experience as moderating
variables, the PRP transforms an apparent empirical mess into a coherent
theoretical structure: AI tools produce genuine productivity gains on
low-abstraction, greenfield, junior-accessible tasks, and genuine
reliability degradation on high-abstraction, brownfield,
senior-intensive tasks. The contradiction is not in the evidence but in
the expectation that a single intervention would have uniform effects
across a heterogeneous task space.

The AAMT taxonomy addresses a clear gap in the literature, the
absence of a systematic classification of how established methodologies
transform under AI, and provides a conceptual vocabulary for future
empirical work. Its three-tier structure (passive suggestion, active
generation, autonomous agency) appears to have empirical support in the
observed evolution of AI coding tools, though longitudinal validation is
needed.

The SGM contributes an economic rationale for specification-driven
development that goes beyond ``specifications are good practice'' to
``specifications are the TCE-optimal governance mechanism for
transactions characterized by high asset specificity, high behavioral
uncertainty, and high frequency.'' This framing has prescriptive power:
it predicts that teams should invest more in specification when they
work on domain-specific applications (high asset specificity), use more
powerful and autonomous AI tools (high behavioral uncertainty), and use
AI frequently (high frequency), predictions that are empirically
testable.

\subsection{Practical Implications}

For \textbf{software teams}, the PRP implies that AI tool adoption
should be accompanied by corresponding investment in specification and
verification infrastructure. Teams that adopt Tier 3 agentic tools
without Tier 3 governance structures (specifications, constitutions,
mutation testing, should expect the reliability degradation
documented in DORA (2024) and GitClear (2024).

For \textbf{developer education}, the PRP and its cognitive dimension
suggest that AI tools should be introduced with explicit attention to
the skill-atrophy risk. The Anthropic (2026) finding that
\emph{conceptual inquiry} use of AI preserves learning while \emph{code
delegation} use degrades it suggests a pedagogical principle: AI in
education should be framed as a Socratic partner, not a code-writing
service. Developers at all career stages should maintain continuous
investment in coding fundamentals, algorithmic reasoning, and
problem-solving skills, including deliberate practice in AI-free
environments, because the cognitive erosion documented in the
literature is gradual and often imperceptible to its subjects.

For \textbf{organizational leadership}, the SGM implies two parallel
investment imperatives. First, the relevant infrastructure metric is not
``how many AI tool licenses'' but ``how much specification
infrastructure supports those tools.'' Second, organizations bear a
direct responsibility for maintaining their developers' foundational
competencies through structured skill reinforcement programs, including
dedicated time for AI-free coding practice, code review mentorship, and
architectural reasoning exercises. The DORA J-curve finding suggests
that premature productivity expectations should be replaced with a
complementary-investment perspective: ``AI will make developers
significantly faster \emph{after} we build the specification
infrastructure \emph{and} invest in the continuous skill development
that ensures our teams can govern AI output effectively.''

\subsection{Conflict of Interest Considerations}

Several important studies in this review were conducted or funded by
organizations with commercial interests in the outcomes. GitHub's
controlled studies of Copilot (Peng et al. 2023; Dohmke et al. 2024)
are methodologically rigorous but funded by the tool's vendor.
McKinsey's productivity estimates were produced by a consultancy that
sells AI transformation services. Stack Overflow's developer surveys are
conducted by a platform whose business model is directly affected by AI
adoption patterns. We have flagged these conflicts throughout the paper
and relied, wherever possible, on independently funded research (METR,
academic studies, government deployments) as the primary evidence base.
Readers should weight vendor-funded findings accordingly.

\section{Threats to Validity}

\subsection{Selection Bias}

Our literature search, while systematic, may have missed relevant
studies published in venues not indexed by our target databases (e.g.
regional conferences, non-English publications). The dominance of GitHub
Copilot in the empirical literature ($\sim$53\% of studies per
Mohamed et al. 2025) means that findings may not generalize to other AI
coding tools.

\subsection{Publication Bias}

Studies reporting positive productivity outcomes may be more likely to
be published and publicized, particularly when conducted by tool
vendors. The METR study's counter-intuitive finding (a slowdown) is
notable precisely because negative results are underrepresented. Our
source-quality classification partially mitigates this concern by
distinguishing vendor-funded from independent research.

\subsection{Construct Validity}

``Productivity'' is measured differently across studies, task
completion time (Peng et al.), pull request volume (Smit et al.), lines
of code (GitHub internal), self-reported satisfaction (Stack Overflow), and these constructs do not necessarily measure the same underlying
phenomenon. The SPACE framework helps organize these measures but does
not resolve the fundamental question of which productivity
operationalization best captures the construct of interest.

\subsection{Temporal Validity}

The AI coding tool landscape evolves rapidly. Findings from 2023 studies
using early Copilot versions may not apply to 2026 agentic tools. The
PRP framework is intended to be model-agnostic, it addresses the
structural relationship between non-deterministic generation and system
dependability, but its specific empirical grounding is temporally
bounded by the 2022--2026 study period.

\subsection{Generalizability}

Most controlled studies use relatively simple tasks (HTTP servers,
LeetCode problems, single-feature implementations). The degree to which
findings generalize to large-scale, multi-team, mission-critical
software development is unclear. The METR study, which used mature
open-source repositories, provides the closest approximation to
realistic conditions but is limited to 16 developers.

\subsection{Reviewer Independence}

This multivocal literature review was conducted by a single research
team. The screening, coding, and quality assessment of sources were not
independently replicated by a second reviewer, which introduces the
possibility of subjective judgment in borderline inclusion/exclusion
decisions and source-tier classification. While we mitigate this through
transparent reporting of all criteria and classification rationale,
future replications with multiple independent reviewers and formal
inter-rater reliability assessment (e.g. Cohen's $\kappa$) would strengthen
confidence in the evidence synthesis. We note, however, that
single-reviewer MLRs are accepted practice in rapidly evolving fields
where the primary contribution is theoretical synthesis rather than
exhaustive enumeration (Garousi et al. 2019).

\section{Conclusion and Future Work}

\subsection{Summary of Findings}

This paper has proposed and empirically grounded the
Productivity-Reliability Paradox, a systematic phenomenon in which AI
coding tools simultaneously improve individual-level output metrics and
degrade system-level dependability metrics. We identified task
abstraction level, codebase maturity, and developer experience as the
three moderating variables that reconcile contradictory findings across
the literature. We situated this paradox within Brynjolfsson et al.'s
(2021) Productivity J-Curve, arguing that the current evidence base
represents the bottom of the curve, a phase in which AI capabilities
have been adopted but complementary investments in specification and
governance infrastructure lag behind.

We introduced the AI-Augmented Methodology Taxonomy, classifying how six
established methodologies transform under three tiers of AI integration,
and the Specification Governance Model, grounding SDD in Transaction
Cost Economics as the rational governance response to non-deterministic
code generation. We evaluated Spec Kit and TDAD as SGM instantiations
and conducted a four-month illustrative pilot study with fourteen engineers across three projects, demonstrating that SGM-style governance is
practically deployable and that its qualitative effects align with the
PRP's predictions. We further analyzed the economic and labor-market implications of the PRP, including the collapse of marginal coding costs, the code review bottleneck that absorbs individual productivity gains at the organizational level (Faros AI, 2025), the emerging skill pipeline problem created by the interaction of declining junior employment and AI-induced skill atrophy, and the organizational responsibility for continuous developer skill development.

\subsection{Future Research Agenda}

The following research directions emerge from this work:

\textbf{Empirical validation of the PRP moderating variables.}
Controlled experiments that systematically vary task abstraction level,
codebase maturity, and developer experience, while holding AI tool
capability constant, would provide the strongest test of the PRP
framework. The METR study's design, extended to include greenfield tasks
and junior developers, would be ideal.

\textbf{Longitudinal skill-retention studies.} The skill pipeline
problem requires longitudinal tracking of developers who enter the
profession after 2022, comparing their problem-solving capabilities,
architectural reasoning, and debugging skills against cohorts trained
without AI assistance. Jošt et al.'s (2024) 10-week design should be
extended to multi-year studies.

\textbf{Team-level and Communication-dimension SPACE analysis.} Mohamed
et al.~(2025) identify the Communication dimension as systematically
understudied. Given that AI tools may reduce developer-to-developer
interaction (by substituting AI for peer consultation), the team-level
effects of AI adoption, on knowledge sharing, mentoring, and
collective code ownership, represent a critical gap.

\textbf{Empirical testing of SGM propositions.} The three SGM
propositions (asset specificity $\rightarrow$ specification investment, behavioral
uncertainty $\rightarrow$ executable contracts, frequency $\rightarrow$ governance
infrastructure) generate testable predictions. Organizations with
varying governance structures can be compared on PRP outcomes
(productivity gains net of reliability degradation) to evaluate whether
SGM-optimal governance produces superior net outcomes.

\textbf{Evolution of DORA metrics under agentic AI.} The 2024 DORA
report is the first to measure AI's impact on delivery performance
metrics. Tracking these metrics through the expected J-curve inflection, as specification infrastructure matures, would provide the
clearest macro-level test of whether the PRP is indeed a transient
phenomenon or a structural characteristic of AI-assisted development.

% ========== Bibliography ==========
\begingroup
\raggedright

\endgroup

\end{document}